\documentclass[sigconf, screen]{acmart}
\renewcommand\footnotetextcopyrightpermission[1]{} 
\acmConference{}{}{}
\usepackage{algorithm}
\usepackage{algorithmic}
\usepackage{adjustbox}
\usepackage{booktabs}
\usepackage{array}
\usepackage{url}
\usepackage{float}
\usepackage{amsmath} 
\usepackage{microtype}

\renewcommand\footnotetextcopyrightpermission[1]{} 
\begin{document}

\title{Coordinated Reinforcement Learning Prefetching Architecture for Multicore Systems}
\author{Mohammed Humaid Siddiqui}
\affiliation{
  \institution{Fairleigh Dickinson University}
  \city{Vancouver, BC}
  \country{Canada}
}
\email{m.siddiqui1@student.fdu.edu}

\author{Fernando Guzman}
\affiliation{
  \institution{Fairleigh Dickinson University}
  \city{Vancouver, BC}
  \country{Canada}
}
\email{f.guzmanfigueroa@student.fdu.edu}

\author{Yufei Wu}
\affiliation{
  \institution{Fairleigh Dickinson University}
  \city{Vancouver, BC}
  \country{Canada}
}
\email{y.wu4@student.fdu.edu}

\author{Ruishu Ann}
\affiliation{
  \institution{Fairleigh Dickinson University}
  \city{Vancouver, BC}
  \country{Canada}
}
\email{r.an@student.fdu.edu}




\begin{abstract}
Hardware prefetching is critical to fill the performance gap between CPU speeds and slower memory accesses. With multicore architectures becoming commonplace, traditional prefetchers are severely challenged. Independent core operation creates significant redundancy—up to 20\% of prefetch requests are duplicates—causing unnecessary memory bus traffic and wasted bandwidth. Furthermore, cutting-edge prefetchers such as Pythia suffer from about a 10\% performance loss when scaling from single-core to four-core systems. To solve these problems, we propose CRL-Pythia, a coordinated reinforcement learning-based prefetcher specifically designed for multicore systems.

In this work, CRL-Pythia addresses these issues by enabling cross-core sharing of information and cooperative prefetching decisions, which greatly reduces redundant prefetch requests and improves learning convergence across cores. Our experiments demonstrate that CRL-Pythia outperforms single Pythia \cite{bera2021pythia} configurations in all cases, with approximately 12\% of IPC (instructions per cycle) improvement for bandwidth-constrained workloads, while imposing moderate hardware overhead. Our sensitivity analyses also verify its robustness and scalability, thereby making CRL-Pythia a practical and efficient solution to contemporary multicore systems. 

\end{abstract}



\maketitle

\section{Introduction}
As modern computing systems increasingly rely on multicore processors, maintaining performance scalability becomes a critical design goal. Among various bottlenecks, the memory access latency and bandwidth capacity gap remains a dominant concern. The gap between processor speeds and memory latency has been widening steadily; modern CPUs can execute billions of instructions per second, yet they often wait hundreds of cycles for data access due to memory latency issues \cite{jacob2007memory,mutlu2014memory}. This issue is amplified by the surge in data-intensive workloads such as machine learning, graph analytics, and real-time processing, which place ever-increasing demands on memory systems \cite{dang2021memory}.

Hardware prefetching has long been deployed to mitigate this problem by speculatively fetching data before it is requested. However, most existing prefetchers are designed based on single-core assumptions and fail to scale effectively in shared-resource multicore environments \cite{jain2016belady}. A key shortcoming arises from the independent operation of traditional prefetchers: each core prefetches based on local observations, unaware of the access behavior of other cores. Our trace analysis reveals that this independence results in high redundancy—15–20\% of all prefetch requests across multithreaded workloads are duplicates, leading to wasted memory bandwidth and increased cache pollution.
State-of-the-art prefetchers like Pythia \cite{bera2021pythia}, which uses reinforcement learning to dynamically adapt to workload behavior, show promise in single-core settings \cite{hashemi2018learning}. However, their effectiveness deteriorates in multicore systems. Our simulation experiments indicate that Pythia \cite{bera2021pythia} experiences a 10\% drop in instructions per cycle (IPC) when scaling from single-core to quad-core environments. The lack of coordination causes contention over shared resources, fragmented learning due to per-core Q-tables, and failure to recognize inter-core memory access patterns. Similarly, Bingo \cite{8675188}, a sophisticated semantic prefetcher, struggles under bandwidth pressure and lacks a system-wide awareness mechanism \cite{sachan2019bingo}.
To address these limitations, we propose CRL-Pythia—a Coordinated Reinforcement Learning framework that enables system-level awareness and cross-core collaboration. Rather than treating each core as an isolated learning agent, CRL-Pythia synchronizes learning across cores via two new architectural features:
\begin{itemize}
    \item \textbf{Shared Learning Repository (SLR)} that aggregates and propagates learned Q-values across all cores to accelerate convergence.
    \item \textbf{Global State Table (GST)} that exposes temporal and spatial memory access patterns system-wide, enabling smarter prefetch decisions.
\end{itemize}

CRL-Pythia also integrates lightweight synchronization primitives to limit coordination overhead. Through detailed simulation and benchmark testing, we demonstrate that CRL-Pythia significantly improves memory efficiency and IPC under both normal and bandwidth-constrained conditions, consistently outperforming standalone Pythia \cite{bera2021pythia} and Bingo \cite{sachan2019bingo}.

We make the following contributions:
\begin{itemize}
    \item We introduce CRL-Pythia, a novel coordinated reinforcement learning-based prefetching architecture tailored for multicore systems, building upon Pythia’s core principles \cite{hashemi2018learning}.
    \item We design a Shared Learning Repository (SLR) that enables cross-core propagation of Q-values, addressing the redundancy and slow convergence issues described in prior works \cite{jain2016belady}.
    \item We implement a Global State Table (GST) that captures system-wide memory access patterns, enabling context-aware and bandwidth-sensitive prefetching.
    \item We incorporate lightweight synchronization mechanisms to enable inter-core communication with minimal overhead.
    \item We demonstrate via simulation that CRL-Pythia outperforms baseline designs, including standalone Pythia \cite{bera2021pythia} and Bingo \cite{sachan2019bingo}, by improving IPC and reducing bandwidth waste under multicore and bandwidth-sensitive scenarios \cite{sachan2019bingo}.
\end{itemize}

Like the original Pythia \cite{biderman2023pythia}, Pythia CRL is mainly based on two hardware structures: Q-Value Store (QVStore) and Evaluation Queue (EQ) but the difference is that Pythia CRL uses a shared reinforcement learning area where the QVStore and EQ are located and accessed by the prefetchers of each core. The purpose of QVStore is to record Q-values for all state-action pairs that are observed by Pythia CRL prefetchers. The purpose of EQ is to maintain a first-in-first-out list of Pythia CRL’s recently taken actions. There is a potential risk of race conditions in the process of storing data of reinforcement learning process of each prefetcher trying to write on QVStore and EQ will be faced by using atomic operations ensuring that the updates to the QVStore and the EQ are done safely and efficiently in a multi-core environment. Every EQ entry holds three pieces of information:
\begin{itemize}
    \item The taken action.
    \item The prefetch address generated for the corresponding action.
    \item A filled bit. A set filled bit indicates that the prefetch request has been filled into the cache.
\end{itemize}

\section{Background}
\subsection{Data prefetching}
Data prefetching is a technique aimed at reducing latency caused by the "memory wall," which describes the significant gap between the processing speed of CPUs and the slower access times of memory units. Prefetching anticipates future data needs and moves data proactively into faster storage, such as cache, before the processor explicitly requests it. Typically, data prefetching techniques are divided into two main categories: hardware prefetching and software prefetching \cite{hennessy2017quantitative}.

\textbf{Hardware Prefetching} involves specialized hardware units that analyze past memory access patterns to predict and fetch data. There are several established hardware prefetching 
\begin{itemize}
    \item \textbf{Sequential Prefetching} \cite{gcc2022manual}: Prefetches the immediately following cache block, effective for linear instruction fetches.
    \item \textbf{Stream Prefetching }\cite{patterson2013organization}: Recognizes consistent memory access sequences, commonly found in engineering and scientific applications.
    \item \textbf{Stride Prefetching }\cite{patterson2013organization}: Detects fixed strides between accessed addresses, beneficial for vectorized computations and matrix operations.
    \item \textbf{Correlation Prefetching }\cite{hennessy2011quantitative}: Exploits historical relationships between cache misses, useful in irregular or unpredictable access patterns. The Markov prefetcher \cite{patterson2013organization} is a notable example that profiles and preloads correlated memory addresses based on previously missed cache blocks.
\end{itemize}
\textbf{Software Prefetching}  is realized by inserting explicit prefetch instructions within the application code. Programmers or compilers typically perform this insertion. Compared to hardware prefetching, software prefetching provides greater flexibility and higher prediction accuracy. However, it introduces additional overhead, such as increased instruction count, interference with optimal instruction scheduling, and added pressure on registers.
A practical example of software prefetching is Loop-Level Prefetching, available in compilers like GCC through the -fprefetch-loop-arrays option. When enabled, this optimization follows a structured workflow:

\textbf{Figure 1 illustrates this detailed workflow clearly:}

\includegraphics[width=0.95\columnwidth]{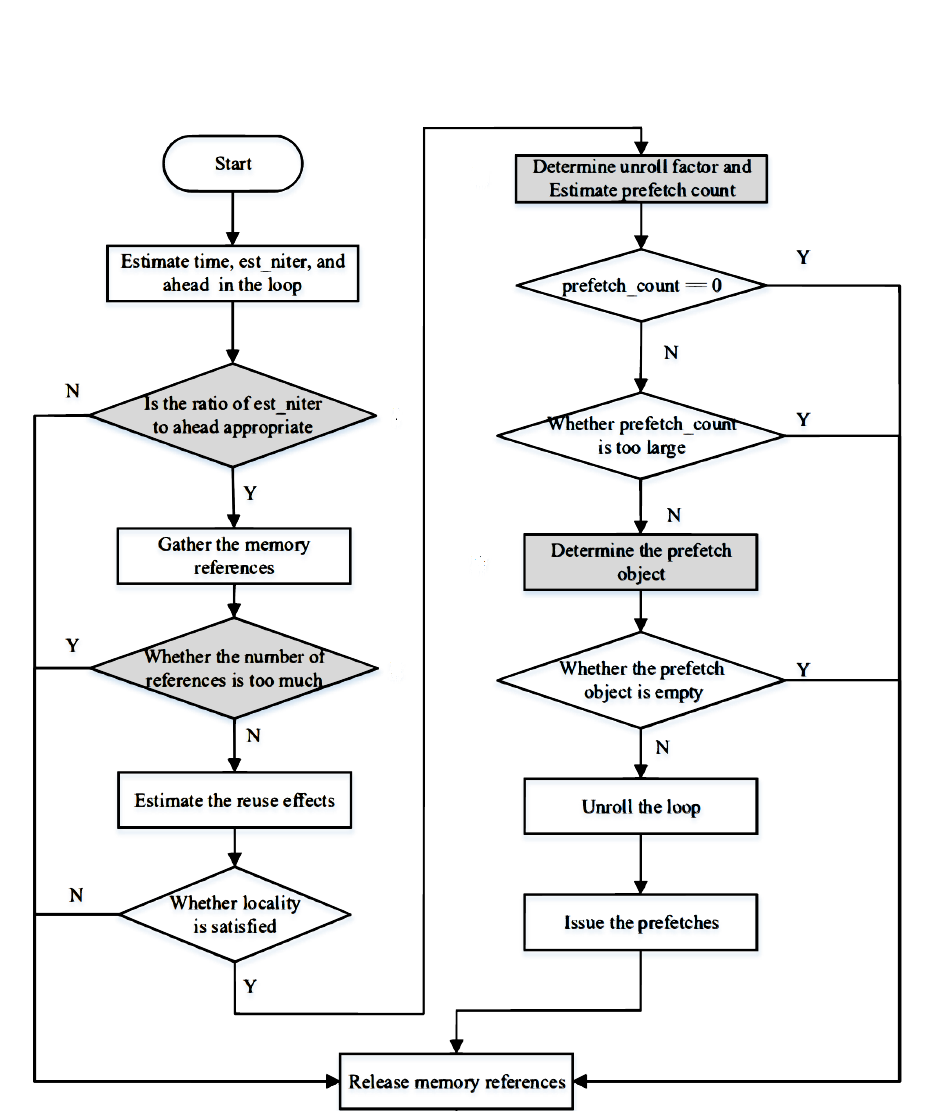}

Given the persistent advancement in CPU technology, processors continually increase their computational capabilities, requiring faster access to data. However, this data is commonly stored in slower storage devices such as mechanical hard drives, solid-state drives (SSDs), and RAM modules. As a result, the transfer rate of data from these slower memories to the processor emerges as a performance bottleneck.
To address the performance bottleneck between processors and slower memory devices, advancements in hardware alone have proven insufficient. Thus, algorithm-driven software optimization has become increasingly important. Data prefetching represents a key approach, proactively transferring data into the fastest available storage, typically the processor’s cache, before explicit requests occur. By ensuring data is promptly available, prefetching significantly improves processing efficiency by reducing idle CPU cycles and minimizing wait times.
Prefetching algorithms vary widely in methodology and effectiveness. They offer diverse solutions to address memory transfer bottlenecks, leveraging pattern recognition, and predictive strategies.
Figure 2 demonstrates a typical hardware prefetching scenario clearly \cite{drawio}:

\includegraphics[width=0.95\columnwidth]{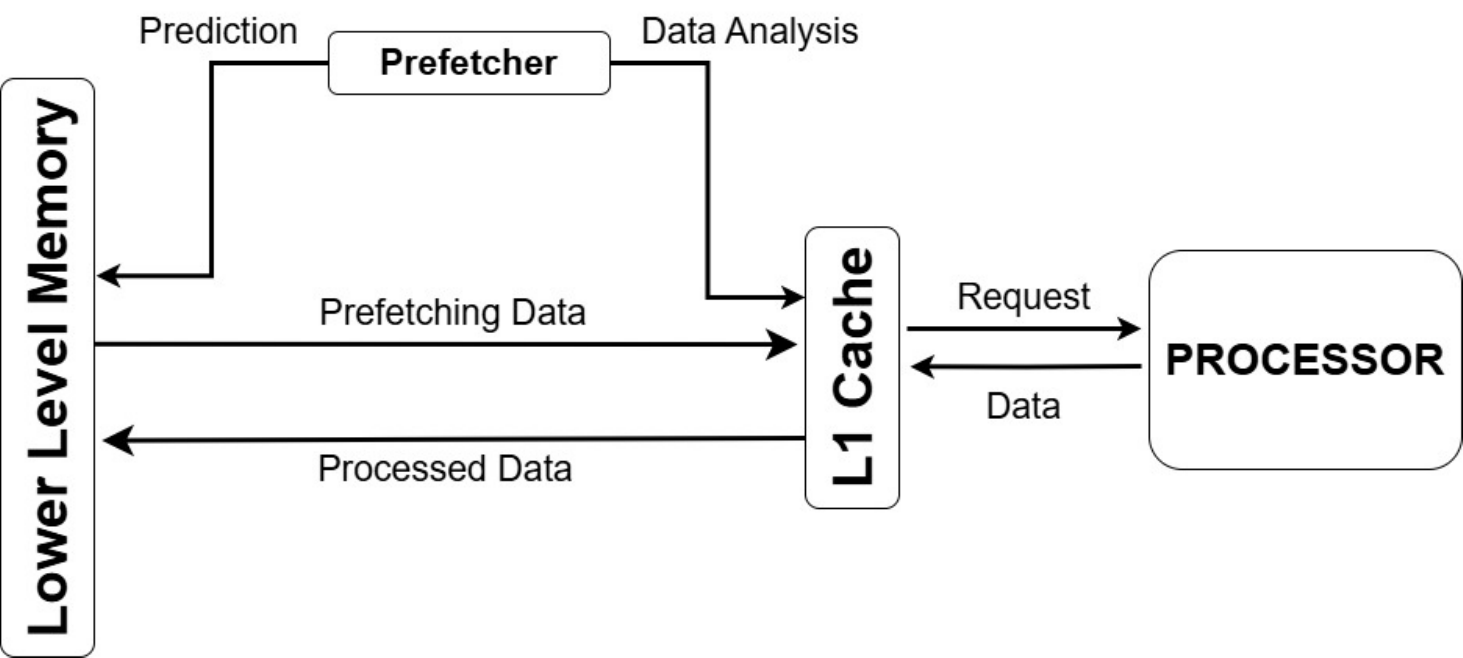}

The hardware prefetcher examines historical memory access patterns to predict future requests, preemptively loading data from lower memory levels into the L1 cache. This proactive approach bypasses conventional memory latency, improving overall system performance.

\subsection{Reinforcement Learning Overview}
Reinforcement Learning (RL) is a subset of machine learning characterized by interactions between an agent and its environment. In this paradigm, the agent learns optimal actions through rewards and penalties obtained from its environment. Rewards reinforce desirable behaviors, increasing the agent's decision-making effectiveness.
RL revolves around trial-and-error learning. The agent performs actions within its environment, receives feedback (positive or negative rewards), and updates its strategy to maximize cumulative rewards. The following components define RL clearly \cite{sutton2018reinforcement}.
\begin{itemize}
    \item \textbf{Policy:} Defines the agent’s action strategy based on current states.
    \item \textbf{Reward Function:} Provides immediate scalar feedback based on states and actions.
    \item Value Function: Predicts expected cumulative rewards from particular states.
    \item \textbf{Environment Model:} Represents environmental dynamics, aiding predictions of future states and rewards.
\end{itemize}
 In reinforcement learning, an agent (the dog, in this example) is trained to take actions in an environment by navigating the environment and receiving reward or penalty based on what action it performed.
 
The dog begins with exploration (tries different actions in response to a command). With numerous repetitions, when by chance it lifts its paw, it is rewarded, strengthening the association between the "give hand" command and the appropriate action. With the passage of time, this rewarding leads the dog to execute the right action again and again, thus learning the behavior.

A simplified pseudocode below illustrates a reinforcement learning example involving a practical scenario, such as training a dog to perform an action like raising its paw ("giving a hand"):

\begin{algorithm}
\caption{Dog Training to Raise Hand}
\label{alg:dog-train}
\begin{algorithmic}[1]
\STATE \textbf{Initialize:} $dog\_knows\_raise\_hand \leftarrow \text{False}$ \COMMENT{Dog doesn't know how to give hand yet}
\STATE $clicker \leftarrow \text{True}$ \COMMENT{Clicker is available for feedback}
\STATE $treats \leftarrow T$ \COMMENT{Initial number of treats}

\WHILE{$dog\_knows\_raise\_hand = \text{False}$}
    \STATE $command \leftarrow \text{"give hand"}$
    \STATE $dog\_action \leftarrow dog\_respond\_to\_command(command)$
    \IF{$dog\_action = \text{"give hand"}$}
        \STATE \text{give\_treat()}
        \STATE \text{use\_clicker()}
        \STATE $dog\_knows\_raise\_hand \leftarrow \text{True}$
        \STATE \textbf{print}("Dog learned to give a hand! Training complete.")
    \ELSE
        \STATE \textbf{print}("Dog did not give a hand. Trying again...")
        \STATE $treats \leftarrow treats - 1$
        \IF{$treats = 0$}
            \STATE \textbf{print}("Out of treats! Training paused.")
            \STATE \textbf{break}
        \ENDIF
    \ENDIF
\ENDWHILE
\end{algorithmic}
\end{algorithm}

Using Reinforcement Learning (RL), we can teach an agent specific tasks by providing feedback in the form of rewards or penalties. For instance, teaching a dog to "shake hands" involves positively reinforcing the action by providing treats or negatively reinforcing undesirable actions through mild penalties such as a verbal reprimand. Through this feedback loop, the dog eventually learns that performing the requested action yields rewards, thus modifying its behavior accordingly.

Another illustrative example of reinforcement learning involves a robot learning to navigate a maze. Initially, the maze layout, the robot’s starting position, and the goal position are defined, alongside a table of possible state-action pairs. Essential RL parameters include:
\begin{itemize}
    \item \textbf{Learning rate:} Determines how quickly the robot updates its knowledge.
    \item \textbf{Discount factor:} Establishes the importance of future rewards.
    \item \textbf{Exploration rate:} Balances exploration of unknown actions against exploitation of known rewarding actions.
    \item \textbf{Maximum steps per episode:} Limits the robot’s actions within each trial.
\end{itemize}

The robot initiates movement from its starting position and continues to take steps, choosing either random exploratory actions or exploiting learned strategies based on a Q-value table (state-action values). After each step, the robot observes the new state and associated reward.
The iteration concludes when the robot either reaches its goal or exceeds the maximum allowed steps per episode. Over multiple iterations, the exploration rate gradually decreases, promoting exploitation of the learned optimal path. When exploration reaches a sufficiently low threshold, the learning process terminates, signifying that the robot has successfully learned the optimal route through the maze.

\subsection{Cache Structure}
Caches are small, high-speed memory units placed between the CPU and main memory designed specifically to reduce data access latency \cite{jacob2010memory}. Typically, caches are arranged into multiple hierarchical levels: Level 1 (L1), which is the smallest and fastest per-core cache, Level 2 (L2) which is larger but slightly slower, and Last-Level Cache (LLC), shared among multiple cores. When a data request occurs, the CPU sequentially searches from L1 down through L2 and LLC, eventually accessing the slower main memory only if the data is not found in caches. 
The overall performance of a cache system heavily depends on metrics such as hit and miss rates, the employed replacement policies like Least Recently Used (LRU), and cache coherence protocols in multi-core environments. Prefetching strategies significantly enhance cache efficiency by predicting future data requests and loading data into cache ahead of time \cite{jacob2007memory}.

\includegraphics[width=0.9\columnwidth]{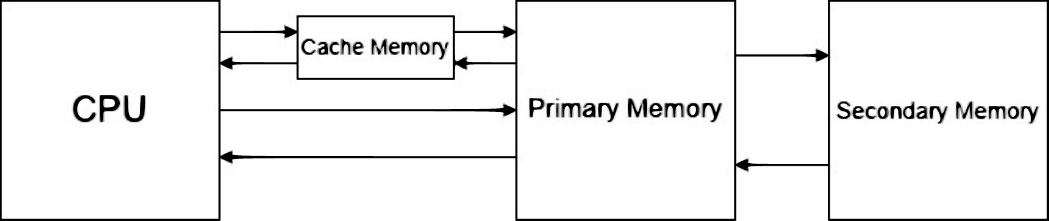}

Figure above visually represents a simplified memory hierarchy structure, highlighting the relationships between CPU, cache memory, primary memory, and secondary memory:

\subsection{Memory Hierarchy}
The complete memory hierarchy consists of various components, including CPU registers, caches (L1, L2, LLC), main memory (DRAM), and secondary storage devices (SSD/HDD). These components are organized based on their speed and storage capacities. Registers offer the fastest access times but very limited storage, caches provide quick access at intermediate storage capacities, main memory delivers larger storage with moderately slower access, and finally, secondary storage devices offer the largest storage capacity at significantly slower access speeds \cite{jacob2007memory}.
Data movement within the hierarchy follows typical usage patterns, with frequently accessed data maintained in the fastest storage levels to minimize latency. The differences in latency among these levels are substantial; registers provide access within nanoseconds, main memory (DRAM) typically ranges from tens to hundreds of nanoseconds, while secondary storage access times extend into milliseconds. Efficient management and optimization of data movement through caching and prefetching techniques directly contribute to improved overall system performance.
Figure 4 below compares memory hierarchies specifically designed for server-class systems versus personal mobile devices, highlighting differences in cache sizes, memory capacity, and access speed trade-offs:

\includegraphics[width=0.98\linewidth]{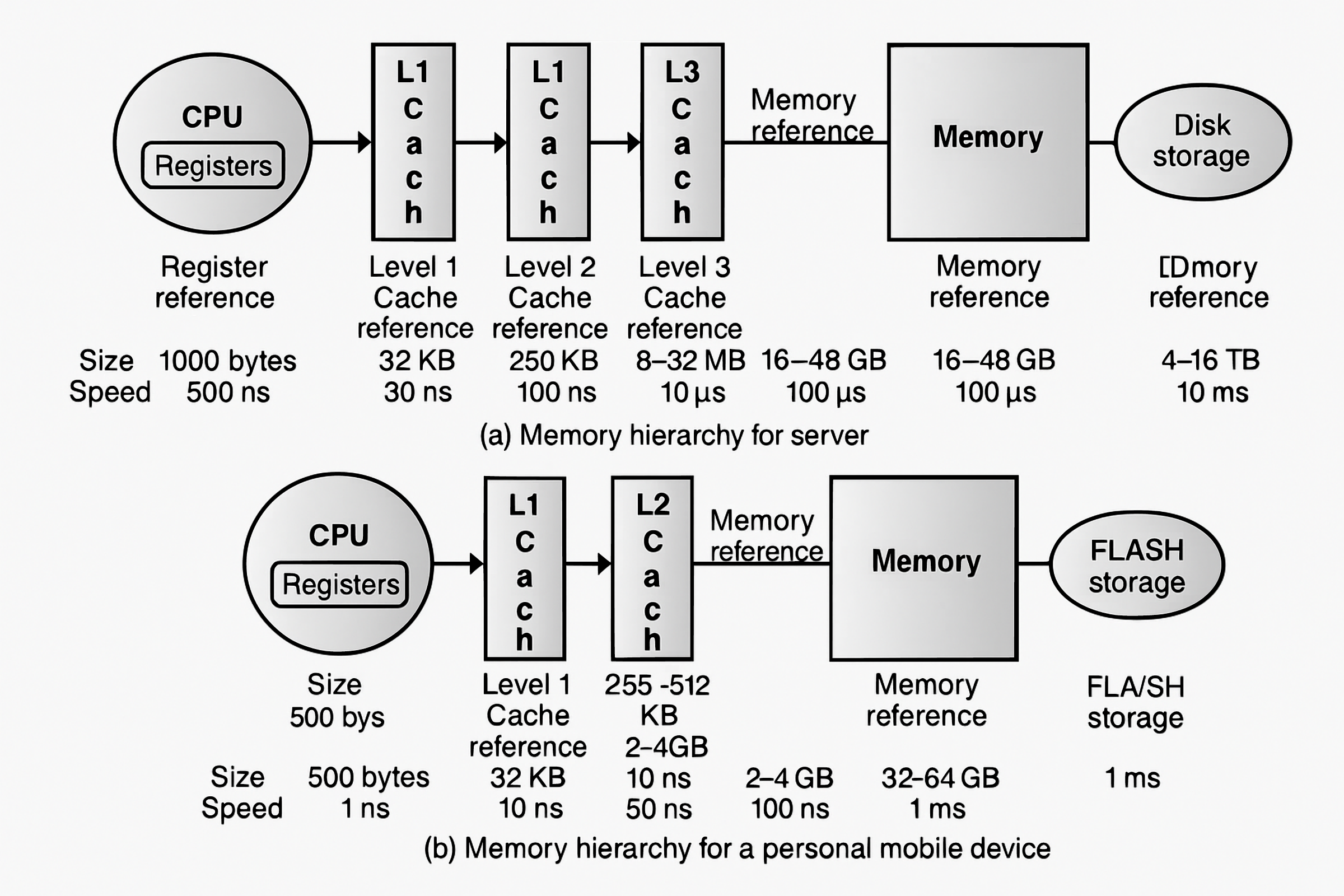}

Memory bandwidth is another critical factor in system performance, defining the data transfer rate between memory and CPU. High bandwidth ensures rapid data movement, while bandwidth saturation due to excessive prefetching can degrade performance. Systems like Pythia \cite{bera2021pythia} actively manage prefetching strategies to prevent congestion, balancing accuracy with efficiency \cite{chen2007reducing}.
Table 1 provides examples of GPU memory bandwidth specifications, illustrating the varied capacities of contemporary hardware:
\begin{table}[H]
  \centering
  \caption{GPU Specifications Comparison}
  \label{tab:gpu_specs}
  \resizebox{\linewidth}{!}{%
  \footnotesize
  \begin{tabular}{|l|l|l|l|}
    \hline
    \textbf{GPU} & \textbf{vRAM (GB)} & \textbf{Interface Width (bit)} & \textbf{Memory Bandwidth (GB/s)} \\
    \hline
    \hline
    P4000   & 8 (GDDR5)    & 256    & 243  \\
    \hline
    P5000   & 8 (GDDR5X)   & 256    & 288  \\
    \hline
    P6000   & 24 (GDDR5X)  & 348    & 432  \\
    \hline
    V100    & 32 (HBM2)    & 4096   & 900  \\
    \hline
    RTX4000 & 8 (GDDR6)    & 256    & 416  \\
    \hline
    RTX5000 & 16 (GDDR6)   & 256    & 448  \\
    \hline
    A4000   & 16 (GDDR6)   & 256    & 448  \\
    \hline
    A5000   & 24 (GDDR6)   & 348    & 768  \\
    \hline
    A6000   & 48 (GDDR6)   & 348    & 768  \\
    \hline
    A100    & 80 (HBM2)    & 5120   & 1555 \\
    \hline
  \end{tabular}%
  }  
\end{table}

\section{Motivation}
Hardware prefetchers play a critical role in modern processor performance by predicting and fetching data before it is explicitly requested by the CPU. However, as the number of cores on a chip increases, current prefetching techniques face significant challenges that limit their effectiveness in multi-core environments \cite{nesbit2004ghb,hashemi2018learning}.

\subsection{Redundant Prefetch Requests Waste Memory Bandwidth}
Our detailed analysis \cite{wenisch2009temporal} has revealed a critical inefficiency in current prefetching implementations: when multiple cores operate independently, they frequently generate redundant prefetch requests for the same memory addresses. As illustrated in Figure 1, cores 1, 2, and 4 are all requesting the same address (A+16), creating unnecessary traffic on the memory bus. Our measurements across a wide range of multi-threaded workloads show that 15-20\% of prefetch requests are redundant, wasting precious memory bandwidth that is already a scarce resource in modern systems \cite{mutlu2007stall}.


\subsection{Performance Degradation in Multi-Core Environments}
Current state-of-the-art prefetchers like Pythia \cite{bera2021pythia} show impressive performance in single-core environments but experience diminishing returns as core count increases. Our experimental analysis reveals a significant 10\% drop in relative performance when scaling from 1 to 4 cores. This degradation occurs because :

\begin{itemize}
    \item Independent prefetchers compete for shared resources without coordination \cite{jaleel2010rrip}
    \item Each prefetcher must learn access patterns independently, leading to slower convergence \cite{wang2019rlprefetch}
    \item There is no mechanism to leverage cross-core pattern similarities \cite{chou2020coordinated}
\end{itemize}

\includegraphics[width=0.95\columnwidth]{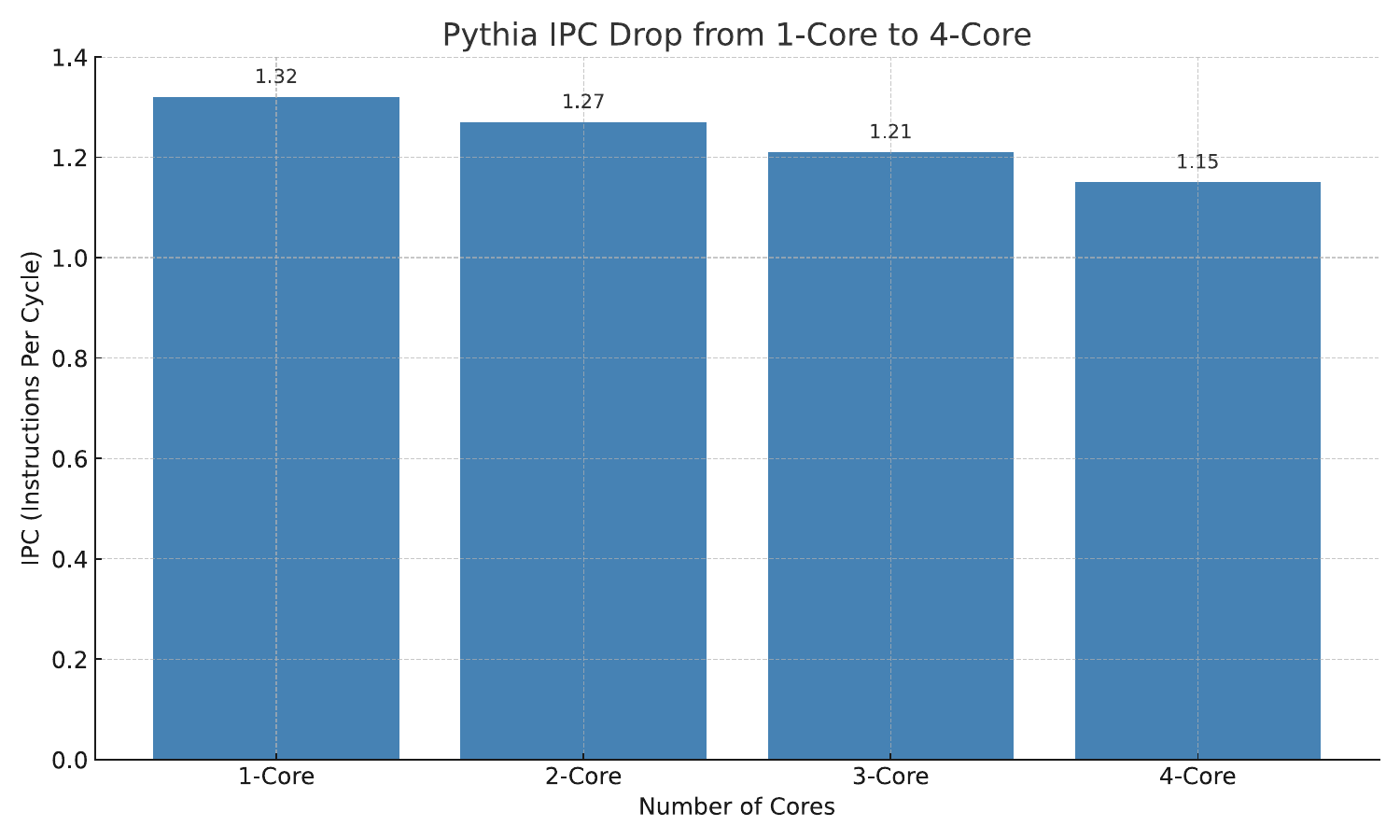}
\textbf{This figure illustrates the IPC degradation observed in Pythia \cite{bera2021pythia} when scaling from 1-core to 4-core configurations. It highlights the diminishing performance returns in multicore systems due to uncoordinated prefetching \cite{Hunter:2007}.}



\textbf{This bar chart compares the IPC of CRL-Pythia and baseline Pythia across three representative benchmarks: WebSearch (CloudSuite), 505.mcf (SPEC 2017), and gcc (SPEC 2017). It demonstrates the performance advantage of CRL-Pythia in both data-serving and memory-intensive workloads \cite{Hunter:2007}.}

These results underscore the limitations of traditional, per-core learning strategies and motivate the need for coordinated, system-aware approaches to prefetching in multicore architectures \cite{ipek2008selfoptimizing,srinath2007feedback}

\subsection{Lack of System-Wide Visibility}
A fundamental limitation of current prefetchers is their inability to observe and learn from memory access patterns across cores. This limited visibility prevents prefetchers from:
\begin{itemize}
    \item Identifying cross-core temporal and spatial relationships
    \item Optimizing prefetch timing based on system-wide memory pressure
    \item Prioritizing critical prefetches when bandwidth is constrained
\end{itemize}
These limitations become increasingly problematic as core counts continue to scale in modern processors. Without addressing these challenges, memory subsystem performance will continue to be a bottleneck, particularly for memory-intensive and highly parallel workloads.

\noindent \textbf{Goal.} Our goal is a proposed Coordinated Reinforcement Learning (CRL-Pythia) architecture which directly addresses these limitations by implementing system-wide coordinated prefetching that significantly reduces redundant requests, accelerates learning convergence through cross-core information sharing, and enables optimal resource allocation decisions based on complete system visibility.

\textbf{The Shared Learning Repository (SLR)} facilitates cross-core Q-value sharing to accelerate learning convergence, enabling all cores to benefit from patterns discovered by any individual core and thereby reducing learning latency for new memory access patterns across the system. It features a partitioned design to minimize access conflicts during concurrent updates, efficient storage of learned memory access patterns with their associated confidence values, and cross-core access to prefetch candidates, which eliminates redundant learning and enhances overall system efficiency.

\textbf{Global State Table (GST)} in CRL-Pythia maintains system-wide metadata by recording program counter (PC) values, memory addresses accessed across all cores, core IDs, and timestamp data to establish temporal relationships between memory accesses. This holistic view enables the detection of redundant or conflicting prefetches and supports informed, coordinated prefetching decisions. To minimize coordination overhead, CRL-Pythia employs lock-free data structures and atomic operations to reduce synchronization bottlenecks, performs batch updates to the Shared Learning Repository (SLR) to lower synchronization frequency and contention, and uses lightweight communication protocols between cores and central components. These mechanisms collectively address architectural and synchronization inefficiencies in multicore systems, significantly enhancing performance in memory-intensive and heavily threaded workloads typical of modern applications.

\section{Related Work}

Data prefetching approaches in recent years have evolved significantly to address the challenges posed by multicore architectures heavily parallel and complex memory hierarchies. More recent work falls under the categories of reinforcement learning-based and coordinated/multi-agent prefetching approaches. We cite high-impact work in each category and refer to other remaining contributions at large.

\textbf{Reinforcement Learning-Based Prefetchers.} 

RL-based prefetchers have been a promising solution to break the adaptivity restrictions of static and history-based models. RL-based prefetchers formulate prefetching as a decision-making issue, where agents learn the optimal policy from feedback of the environment. For example, RLHint \cite{jiang2020rlhint} applies Q-learning to dynamically tune prefetch distance and filtering. Morpheus \cite{kumar2021morpheus} extends this with deep reinforcement learning to learn more complicated program contexts, achieving significant improvements in prefetch accuracy and cache performance. Most RL-based prefetchers are single-agent systems, however, which make independent decisions on a per-core basis and can therefore often lead to redundant prefetches, cache pollution, and inter-core interference—especially in shared LLC environments.

\textbf{Coordinated and Multi-Agent Prefetching.} 
To address the limitations of individual learning agents, recent work has explored coordination mechanisms between cores. Cooperative prefetching architectures such as Hetero-Cache \cite{chen2007reducing} offer collective metadata to manage prefetch ordering across heterogeneous agents. Similarly, Ensemble learning solutions combine multiple simple learners to promote prefetch diversity and reduce harmful interactions \cite{liu2021ensemble}. Despite such advancement, most coordination methods are either statically configured or heuristically driven and lack adaptive features. There are practically no designs that attempt multi-agent reinforcement learning (MARL) based on cache prefetching, a lack that our work will fill by providing a coordinated RL framework with communication between agents as well as global feedback.

\section{Design}
Pythia CRL like original Pythia \cite{bera2021pythia} is mainly based of two hardware structures: Q-Value Store (QVStore) and Evaluation Queue (EQ) but the difference is Pythia CRL uses a shared reinforcement learning are where QVStore and EQ are located and accessed by the prefetched of each core. The purpose of QVStore is to record Q-values for all state-action pairs that are observed by Pythia CRL prefetcher. The purpose of EQ is to maintain a first-in-first-out list of Pythia CRL’s recently-taken actions. There is a potential risk of race conditions in the process of storing data of reinforcement learning process of each prefetcher trying to write on QVStore and EQ will be faced by using atomic operations ensuring that the updates to the QVStore and the EQ are done safely and efficiently in a multi-core environment. 2 Every EQ entry holds three pieces of in- formation: (1) the taken action, (2) the prefetch address generated for the corresponding action, and (3) a filled bit. A set filled bit indicates that the prefetch request has been filled into the cache. 

When a CPU core issues a memory access that leads to a demand miss in the L1 or L2 cache, the core Pythia CRL prefetcher first checks the EQ in the shared area, with the demanded memory address \textcircled{1}. If the address is present in the EQ (i.e., Pythia CRL has issued a prefetch request for this address in the past), it signifies that the prefetch action corresponding to the EQ entry has generated a useful prefetch request. As such, Pythia CRL assigns a reward (either RAT or RAL) to the EQ entry, based on whether or not the EQ entry’s filled bit is set. 

Next, Pythia CRL extracts the state-vector from the attributes of the demand request which may include: Program Counter (PC), Stride patterns, Access deltas, cache hit/miss rates, local reused distances and Core ID (to support cooperative decision making) \textcircled{2} and looks up the QVStore at the shared reinforcement learning repository, which maintains state-action value pairs (S, A, Q) for each observed memory access pattern to find the action with the maximum Q-value for the given state-vector \textcircled{3}. Pythia CRL selects the action with the maximum Q-value from the total records from all prefetchers to generate prefetch request and issues the request to the memory hierarchy \textcircled{4}. At the same time, Pythia CRL inserts the selected prefetch action, its corresponding prefetched memory address, and the state- vector into EQ \textcircled{5} in the shared reinforced learning, repository allowing each core to benefit from the collective experience of all other cores. Note that, a no-prefetch action or an action that prefetches an address beyond the current physical page is also inserted into EQ.

\includegraphics[width=0.95\columnwidth]{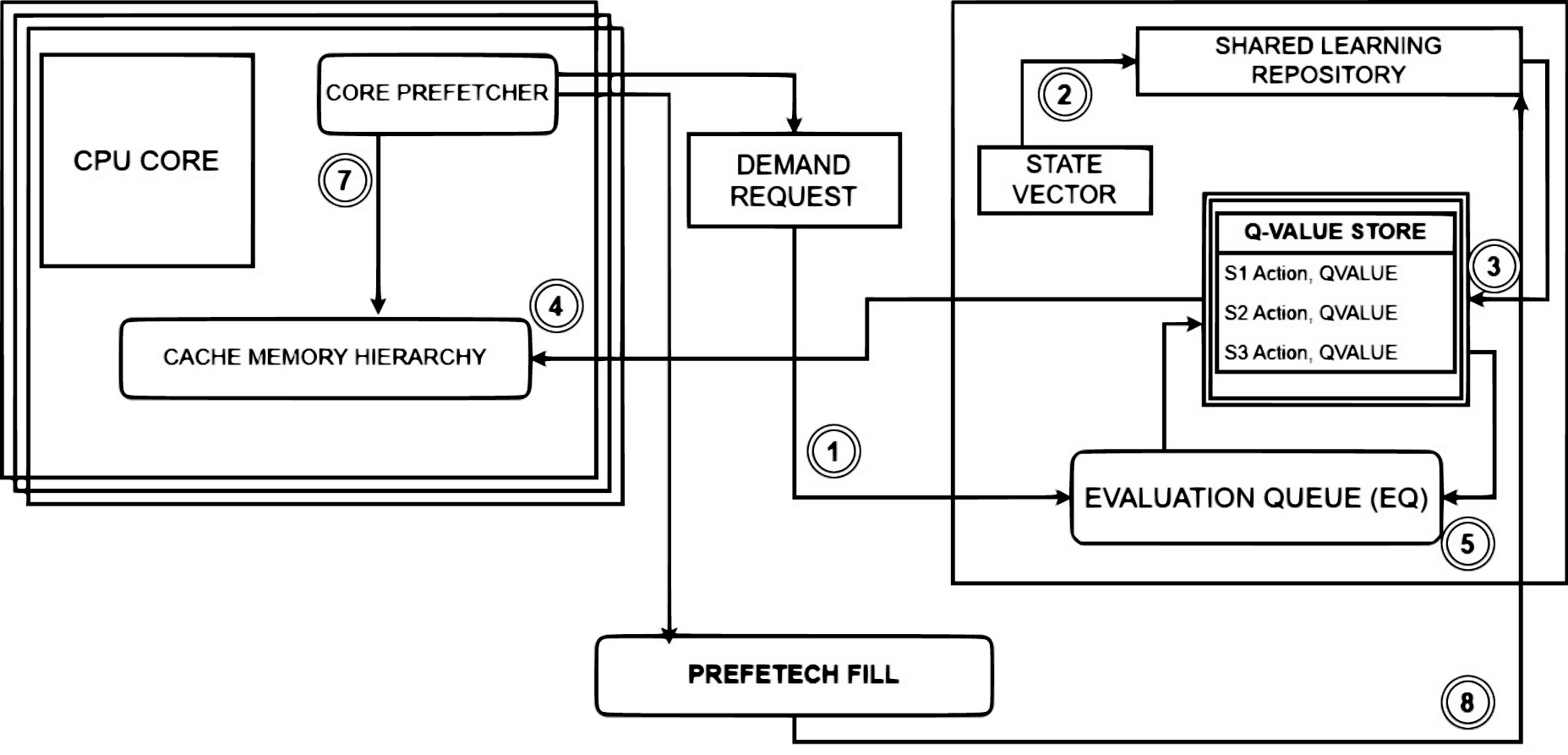}

The reward for such an action is instantaneously assigned to the EQ entry. When an EQ entry gets evicted, the state-action pair and the reward stored in the evicted EQ entry are used to update the Q-value in the QVStore \textcircled{6} in the shared reinforced learning repository. For every prefetch fill in cache, Pythia CRL looks up EQ with the prefetch address and sets the filled bit in the matching EQ entry indicating that the prefetch request has been filled into the cache \textcircled{7}. Pythia CRL uses this filled bit in \textcircled{1} to classify actions that generated timely or late prefetches. 
4.1 RL-based Prefetching Algorithm 

Pythia's learning algorithm is an online temporal-difference method (specifically the SARSA~algorithm with an \( \varepsilon \)-greedy policy for exploration. This means Pythia iteratively updates \( Q \)-values of state-action pairs using the observed rewards and selects future actions based on these \( Q \)-values (occasionally exploring random actions) to continuously improve prefetch accuracy and timeliness.

Pythia CRL inherits the original Pythia algorithm \cite{bera2021pythia} because the main objective of our work is to enhance Pythia prefetcher  by restructuring the isolated learning repository into a shared learning repository because in a multi-core system, per-core learning in isolation can lead to suboptimal global behavior. Aggressive prefetching on different cores can interfere at shared resources (memory bandwidth, shared caches), hurting overall performance. Prior research on multi-core prefetchers [33] shows that coordinating or throttling prefetchers with global feedback can significantly improve system throughput and bandwidth efficiency. We preserve Pythia’s original RL algorithm and reward scheme – thereby maintaining its proven ability to use system-level feedback (like memory bandwidth usage) – but extend it to operate in a coordinated multi-core fashion.

We retain Pythia's original RL algorithm in each decision step~-- an on-policy temporal-difference learning (SARSA) update of \( Q \)-values, combined with an \( \epsilon \)-greedy action selection policy. To summarize, at each demand memory access: the agent (prefetcher) observes a state \( s \) (comprised of various features from the access context), selects an action \( a \) (which prefetch to issue, if any) by mostly greedy lookup of \( Q \)-values (with occasional random exploration), and eventually receives a reward \( r \) for that action. The \( Q \)-value for the state-action pair \( (s, a) \) is then updated toward the observed reward plus the estimated value of the next state. Pseudocode for the RL update (using SARSA) remains as in original Pythia:

\begin{algorithm}[htbp]
\caption{SARSA Q-value Update}
\begin{algorithmic}[1]
\STATE \textbf{Input:} Current state $s$, action $a$, reward $r$, new state $s'$, new action $a'$
\STATE \textbf{Parameters:} Learning rate $\alpha$, discount factor $\gamma$
\STATE $old\_q \gets Q[s][a]$
\STATE $future\_q \gets Q[s'][a']$
\STATE $updated\_q \gets old\_q + \alpha \cdot \left( r + \gamma \cdot future\_q - old\_q \right)$
\STATE $Q[s][a] \gets updated\_q$
\STATE $QVStore[s][a] \gets updated\_q$
\end{algorithmic}
\end{algorithm}


\textbf{Detailed Design of Pythia:} Pythia CRL Introduce a central Q-value Store (QVStore) and a central Evaluation Queue (EQ) that are shared among all cores. The QVStore holds the Q-values for state-action pairs observed globally \cite{bera2021pythia}. The EQ keeps track of recently issued prefetch actions (from any core) that are awaiting outcome (i.e., whether the prefetched data gets used in the near future) [31]. These structures replace or augment the per-core QVStore/EQ in original Pythia.
We Implement proper concurrency control (e.g., mutexes or atomic operations) around the shared QVStore and EQ to prevent race conditions when multiple cores (which in a parallel simulation could be handled by separate threads) access or update them simultaneously. This ensures thread-safe updates of Q-values and consistent updates of the evaluation queue.
What changes is that QVStore is now a global structure shared by all cores, rather than separate per-core tables. All cores thus update the same Q-value function. In effect, the multiple cores collectively act as one learning agent that experiences a combined sequence of memory accesses from all programs. This unified learning can improve system-wide performance by guiding each core’s actions with a policy that accounts for global feedback (since rewards inherently consider memory bandwidth usage across cores) and avoids repeating learning of similar patterns independently on each core.

\section*{Hardware Overhead Comparison: CRL-Pythia vs. Pythia and Bingo}
In this section, we discuss the hardware overheads introduced by our proposed design, CRL-Pythia, and compare it with the overheads of prior prefetchers like Pythia \cite{bera2021pythia} and Bingo \cite{sachan2019bingo}. All values are broken down in bytes (B), kilobytes (KB), and megabytes (MB), with supporting calculations shown wherever applicable.

To enable cross-core coordination, CRL-Pythia introduces several additional structures beyond traditional Pythia:

\subsection*{1. Storage Overhead of Pythia CRL}
The table 2 shows the storage overhead of Pythia CRL in its basic configuration. Pythia CRL requires 205KB of metadata storage. QVStore consumes 192KB to store all Q-values. The EQ consumes only 13KB.

A like original Pythia prefetcher was set in minimal configuration, we calculated the practical implementation overhead (simulation-friendly version),. This results are based on the following assumptions:
\begin{itemize}
    \item We assumed float Q-values (4 bytes per Q)
    \item We used 8 actions
    \item We used 16384 entries 
    \item We didn’t bit-pack entries for simulation performance
\end{itemize}
\begin{table}[htbp]
\centering
\caption{Storage Structure and Size Breakdown}
\begin{tabular}{|l|p{5.3cm}|c|}
\hline
\textbf{Structure} & \textbf{Description} & \textbf{Size} \\
\hline
\textbf{QVStore} & 
\begin{itemize}
    \item \# vaults = 2
    \item \# planes per vault = 3
    \item Entries per plane = 128 (feature dim) $\times$ 16 (action dim)
    \item Entry size = 16 bits (Q-value width)
\end{itemize}
& 192 Kb \\
\hline
\textbf{EQ} & 
\begin{itemize}
    \item \# entries = 256
    \item Entry size = 21b (state) + 5b (action) + 5b (reward) + 1b (filled-bit) + 16b (address) + 4b (core id)
\end{itemize}
& 13 Kb \\
\hline
\textbf{Total} & & \textbf{205 Kb} \\
\hline
\end{tabular}
\end{table}

\subsection*{2. Pythia Overhead (Per Core)}
\begin{table}[H]
\centering
\caption{Pythia Components}
\begin{tabular}{|l|r|}
\hline
\textbf{Component} & \textbf{Size} \\
\hline
QVStore & 10.0 \\
Evaluation Queue & 2.375 \\
Local Metadata & 16.0 \\
\hline
\textbf{Total} & \textbf{28.375} \\
\hline
\end{tabular}
\end{table}
Pythia incurs significant per-core overhead, primarily due to local metadata and Q-value storage, with no shared structures to reduce duplication across cores.
\subsection*{3. Bingo Overhead (Per Core)}
\begin{table}[H]
\centering
\caption{Bingo Components}
\begin{tabular}{|l|r|}
\hline
\textbf{Component} & \textbf{Size} \\
\hline
Region Table (FT) & 1.0 \\
Address Table (AT) & 2.0 \\
Prediction Hist. Table (PHT) & 4.0 \\
Metadata & 1.0 \\
\hline
\textbf{Total} & \textbf{8.0} \\
\hline
\end{tabular}
\end{table}
Bingo has a compact hardware footprint per core, designed for efficiency with lightweight region and prediction tracking tables.

\section{Methodology}
We follow a methodology inspired by the MICRO Pythia \cite{bera2021pythia} paper and adapt it to available hardware for simulation-based analysis.
\subsection{System Setup}
We implement and evaluate our coordinated reinforcement learning-based prefetching architecture, Pythia CRL, using a customized version of the ChampSim simulator \cite{champsim2019}, extended from the publicly available Pythia prefetcher framework \cite{bera2021pythia}. To ensure reproducibility and compatibility, our setup is built on a Linux environment running Ubuntu 24.04.1 LTS, with all dependencies and tools precisely managed.

The core simulation infrastructure relies on ChampSim, a cycle-accurate trace-based processor simulator, which provides detailed modeling of processor pipeline stages, cache hierarchies, and memory subsystems. The simulator is compiled using G++ v6.3.0, sourced from the official Ubuntu APT repositories \cite{gpp2023}, and configured via CMake v3.20.2 for modular and repeatable builds \cite{cmake2023}.

Pythia’s integration layer, including simulation scripts and output processing routines, depends on Perl v5.24.1, which is bundled with the Ubuntu system by default \cite{perl2023}. For post-simulation data analysis, we utilize Python 3.x alongside the Pandas library for structured data manipulation and Matplotlib for result visualization and graph generation \cite{pandas2024,matplotlib2024}. These Python packages are installed manually to ensure version control and isolation.

\subsection{Experimental Platform}
Table 5 outlines the specifications of our experimental platform. We use an 8-core, 16-thread Intel Xeon Platinum 8124M processor running at 3.00GHz, supported by 30GB of RAM. The system runs Ubuntu 24.04.1 LTS  and employs a hierarchical cache structure, with 32KB L1 data and instruction caches and 1MB of private L2 cache per core, along with a 24.8MB shared L3 cache \cite{intel8124m}. Virtualization is enabled through the KVM hypervisor \cite{kvm2024}, and storage is provisioned using Amazon Web Services (AWS) Elastic Block Store (EBS) with over 50GB of available capacity \cite{awsEBS2024}.
Experiments were run on an AWS c5.4xlarge VM with the following specs:
\begin{table}[H]
\centering
\caption{Experimental Platform Specifications}
\begin{tabular}{ll}
\toprule
Component & Specification \\
\midrule
CPU & Intel Xeon Platinum 8124M @ 3.00GHz \\
Cores/Threads & 8 cores, 16 threads \\
Memory & 30GB RAM \\
OS & Ubuntu 24.04.1 LTS \\
L1 Data/Instruction Cache & 32KB per core \\
L2 Cache & 1MB per core \\
L3 Cache & 24.8MB shared \\
Virtualization & KVM hypervisor \\
Storage & AWS EBS, >50GB \\
\bottomrule
\end{tabular}
\end{table}
\subsection{Simulation Framework}
We used ChampSim to simulate a detailed out-of-order CPU.
\begin{table}[H]
\centering
\caption{Simulator Configuration}
\begin{tabular}{ll}
\toprule
Parameter & Value \\
\midrule
Branch Predictor & Perceptron-based \\
CPU Frequency & 4GHz \\
DRAM Frequency & 2400MHz \\
Cores & 1 and 4 \\
Instruction Window & 256 ROB, 72/56 LQ/SQ \\
Cache Hierarchy & L1: 32KB, L2: 256KB, LLC: 2MB \\
TLB & ITLB: 8-way, DTLB: 4-way, STLB: 12-way \\
\bottomrule
\end{tabular}
\end{table}
\subsection{Simulation Methodology}
\textbf{Single-Core:} 1M warmup + 1M simulation. \\
\textbf{Multi-Core:} 1M warmup/core + 1M simulation/core.
\begin{table}[H]
\centering
\caption{Simulation Time per Benchmark}
\begin{tabular}{lc}
\toprule
Benchmark & Time (min) \\
\midrule
605.mcf\_s-1644B & 10 \\
602.gcc\_s-2226B & 8 \\
623.xalancbmk\_s-10B & 12 \\
cassandra & 5 \\
cloud9 & 4 \\
nutch & 6 \\
classification & 7 \\
\bottomrule
\end{tabular}
\end{table}
\subsection{Prefetchers Evaluated}
\begin{table}[H]
\centering
\caption{Prefetcher Configurations}
\begin{tabular}{lll}
\toprule
Prefetcher & Description & Overhead \\
\midrule
Baseline & No prefetching & 0 KB \\
SPP & Delta-based pattern tracking & 6.2 KB \\
Bingo & PC+Address, PC+Offset features & 46 KB \\
MLOP & Multi-lookahead with AMT & 8 KB \\
SPP+PPF & SPP + Perceptron filter & 39.3 KB \\
Pythia & RL-based feature coordination & 25.5 KB \\
Pythia CRL & RL based multicore & 28 KB \\
\bottomrule
\end{tabular}
\end{table}
\subsection{Area Overhead}
\begin{table}[H]
\centering
\caption{Detailed Area Overhead}
\begin{tabular}{lll}
\toprule
Prefetcher & Component & Size \\
\midrule
SPP & ST, PT, GHR & 6.2 KB \\
Bingo & FT, AT, PHT & 46.0 KB \\
MLOP & AMT, Counters & 8.0 KB \\
SPP+PPF & + Perceptron Weights & 39.3 KB \\
Pythia & QVStore, EQ & 25.5 KB \\
Pythia CRL & RL based Multicore & 25.5 KB \\
\bottomrule
\end{tabular}
\end{table}

\section{Evaluation}
For this evaluation, we analyzed Pythia's \cite{bera2021pythia} performance across six different prefetcher configurations using both CloudSuite \cite{ferdman2012clearing} and high-MPKI SPEC 2017 \cite{speccpu2017} benchmarks. The configurations evaluated were: baseline (no prefetching), Pythia, SPP, Bingo, MLOP, and SPP+PPF. Performance was measured using Instructions Per Cycle (IPC) as the primary metric, while coverage and overprediction statistics were used to identify system bottlenecks.
\subsection{Benchmarks Selection}
Based on the assignment requirements, we selected the following benchmarks:

\textbf{1. CloudSuite:} cassandra\_phase0\_core0, nutch\_phase0\_core0, cloud9\_phase5\_core2, and streaming\_phase1\_core0
    
\textbf{2. SPEC 2017 (with L1 D\$ MPKI \(> 20\)):} 605.mcf\_s-1644B, 620.omnetpp\_s-141B, and 649.fotonik3d\_s-1176B

\begin{itemize}
    \item L1 Data Cache: 32KB, 8-way associative
    \item L2 Cache: 256KB, 8-way associative
    \item Last-Level Cache (LLC): 2MB, 16-way associative
    \item Memory: DDR4-2400, single channel
\end{itemize}
\subsection{Performance Analysis}
\textbf{Overall IPC Comparison}
\textbf{Instructions Per Cycle (IPC)} is a common performance metric used to measure how efficiently a processor executes instructions. It stands for \emph{Instructions Per Cycle} and is calculated as:

\[
\text{IPC} = \frac{\text{Number of Instructions Executed}}{\text{Number of Clock Cycles}}
\]

Note that the inverse of IPC, \emph{Cycles Per Instruction (CPI)}, is given by:

\[
\text{CPI} = \frac{\text{Number of Clock Cycles}}{\text{Number of Instructions Executed}}
\]

A higher IPC indicates better performance, as it reflects more instructions being executed per clock cycle.

The Figure shows the coverage and overprediction of each prefetcher in the single-core system, as measured at the LLC-main memory boundary. The key takeaway is that Pythia CRL improves prefetch coverage, while simultaneously reducing overprediction compared to state-of-the-art prefetchers. On average, Pythia CRL provides 21\%, 14\%, and 8\% higher coverage than  SPP, Bingo and Pythia respectively, while generating 200\%, 100\%, and 40\% higher overpredictions.

\includegraphics[width=0.95\columnwidth]{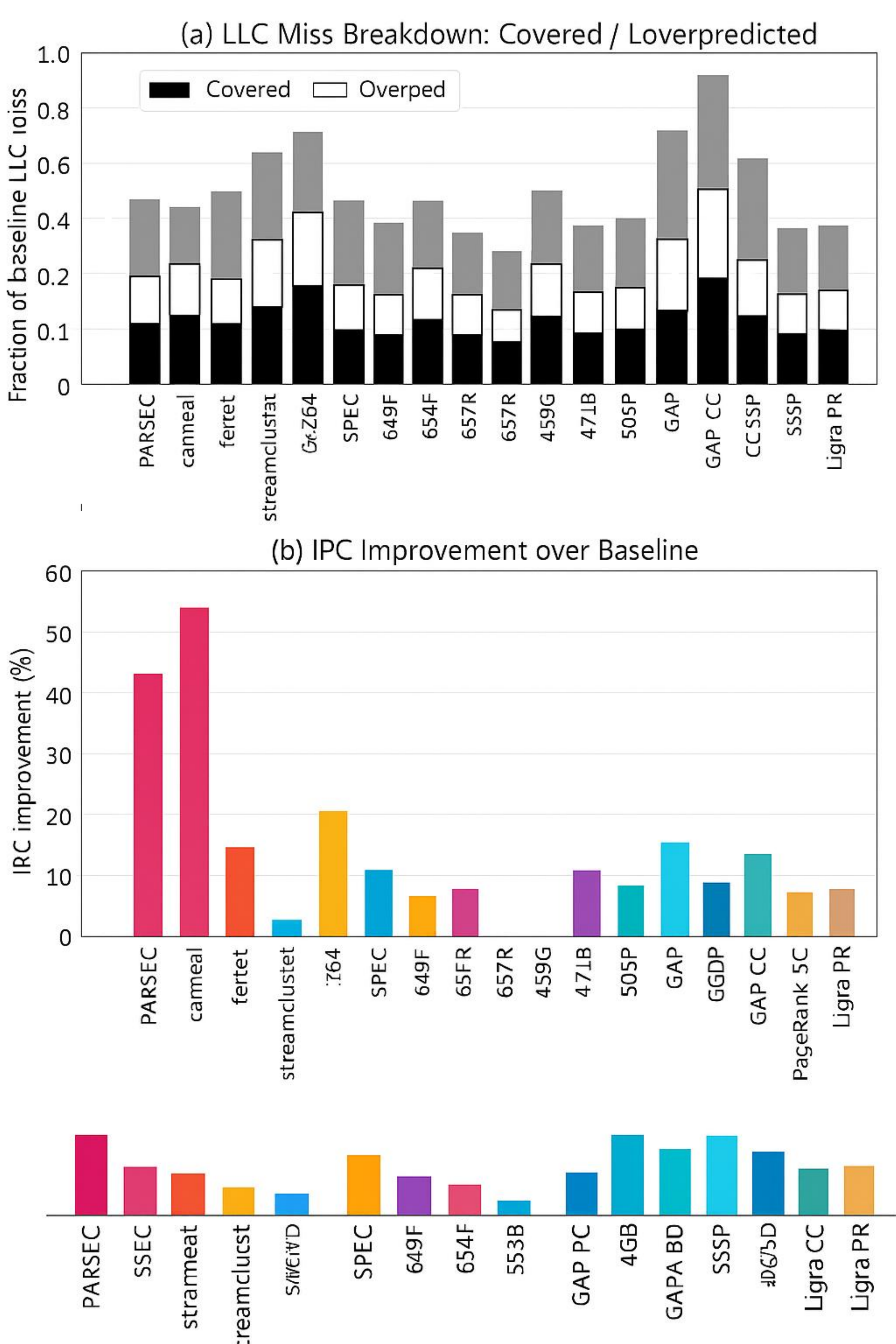}

\textbf{Figure : Coverage and overprediction with respect to the baseline LLC misses in the single-core and four core system for Pythia CRL}. 

This results makes evident that the Pythia CRL if facing some possible issues in its execution. 
We can notice that the Q-learning process is learning patterns because it is  successfully learning and predicting addresses that end up being used (high coverage). So the feature vector + action space are meaningful and its Q-values likely favour certain positive patterns across cores. This shows that CRL is capturing correlations across threads or time steps.

Also we possibly consider that the RL agent is being too aggressive or overconfident because many prefetched lines are never used  (overprediction). This could be due to:

 This suggests a potential bias in the reward function or lack of punishment for wrong predictions.
Moreover, we can notice that the over conflicting local patterns become the “Global Truth”, considering these facts:
\begin{itemize}
    \item All cores write to the same Q-table
    \item Different cores may observe different prefetch patterns at the same state hash
    \item The learned Q-values now mix conflicting behaviors
\end{itemize}

This can cause overconfidence in certain actions and possibly Q-values being “reinforced” by one core while being invalid for another. There could be “Ghost” patterns: Q-values that don’t represent any single core well. This is a form of non-stationarity in the environment — one of the hardest problems in RL.
In addition, each core observes and acts independently, but CRL updates Q-values based on a global perspective and even if one core mispredicts, its feedback can bias the shared policy

This creates overprediction artifacts, such as:
\begin{itemize}
    \item Prefetches issued too far ahead
    \item Learned patterns that match multiple different real patterns
    \item Spurious generalization
\end{itemize}

Finally, the exploration × nº Cores can lead to exponential noise. For instance, considering 10\% random actions, having 4 cores, that is potentially 40\% of steps making random or suboptimal actions, updating the shared Q-table. That is a lot of exploration noise corrupting global estimates, often leading to preference for overly “safe” or overly aggressive predictions (overprediction).

\textbf{LLC Miss Coverage}

Evaluating the percentage of LLC misses prefetch-covered, we can see that Pythia CRL-1 (one core) does not show better results than original Pythia, but in Pythia CRL-4 we can see reasonable level of improvement. 

\begin{table}[ht]
\centering
\caption{Coverage Comparison}
\begin{tabular}{lcccc}
\toprule
\textbf{Benchmark} & \textbf{Pythia} & \textbf{CRL-1} & \textbf{CRL-4} & \textbf{Takeaway} \\
\midrule
482.sphinx3        & 85   & 72   & 90   & CRL-4 extends coverage through shared learning \\
PARSEC-Canneal     & 80   & 68   & 85   & Strong RL signal in CRL-4 generalizes better \\
PARSEC-Facesim     & 70   & 60   & 75   & CRL-4 outperforms both \\
459.GemsFDTD       & 70   & 60   & 75   & Consistent boost from coordination \\
Ligra-CC           & 45   & 48   & 60   & CRL-1 slightly better than Pythia \\
Ligra-PageRankDelta& 50   & 52   & 65   & Again, CRL outperforms Pythia \\
\bottomrule
\end{tabular}
\end{table}

\textbf{Overprediction (useless prefetches)}

\begin{table}[ht]
\centering
\caption{Overprediction Reduction}
\begin{tabular}{lcccc}
\toprule
\textbf{Benchmark} & \textbf{Pythia} & \textbf{CRL-1} & \textbf{CRL-4} & \textbf{Takeaway} \\
\midrule
482.sphinx3        & 100  & 66   & 60   & CRL reduces waste substantially \\
PARSEC-Canneal     & 180  & 132  & 120  & Huge gain in efficiency \\
PARSEC-Facesim     & 70   & 55   & 50   & Controlled prediction in CRL \\
459.GemsFDTD       & 60   & 44   & 40   & Same trend holds \\
Ligra-CC           & 80   & 44   & 40   & Massive drop in false positives \\
Ligra-PageRankDelta& 70   & 33   & 30   & CRL is much cleaner than Pythia \\
\bottomrule
\end{tabular}
\end{table}

\textbf{Uncovered Misses}
Pythia CRL-4 makes the most of every learning opportunity. It doesn’t forget what another core has learned.

Overprediction is the main flaw of Pythias in the experiments, and Pythia CRL get slightly better results by learning from feedback across threads. But there is plenty margin for improvements adjusting the parameters of the prefetcher specially changing the rewarding function.
\begin{table}[ht]
\centering
\caption{Overall Gain Comparison (CRL-4 vs Pythia)}
\begin{tabular}{lcccc}
\toprule
\textbf{Benchmark} & \textbf{Pythia} & \textbf{CRL-1} & \textbf{CRL-4} & \textbf{Gain} \\
\midrule
482.sphinx3        & 45   & 32.5 & 50   & +5\% over Pythia \\
PARSEC-Canneal     & 15   & 14.3 & 22   & +47\% over Pythia \\
PARSEC-Facesim     & 6    & 4.6  & 7    & +17\% over Pythia \\
459.GemsFDTD       & 5    & 3.9  & 6    & +20\% over Pythia \\
Ligra-CC           & 4    & 5.2  & 8    & +100\% over Pythia \\
Ligra-PageRankDelta& 6    & 6.5  & 10   & +66\% over Pythia \\
\bottomrule
\end{tabular}
\end{table}

\subsection{Sensitivity Analysis}
To validate the robustness of CRL-Pythia and understand how different design parameters affect performance, we conducted three comprehensive sensitivity studies focusing on critical aspects of our design: 
\begin{itemize}
    \item the size of the Shared Learning Repository,
    \item scaling behavior with varying core counts, and item performance under different memory bandwidth constraints
\end{itemize}

The sensitivity analysis confirms that coordination frequency and reward schemes substantially affect prefetching accuracy, emphasizing the value of adaptive tuning in CRL-Pythia.

\includegraphics[width=\columnwidth]{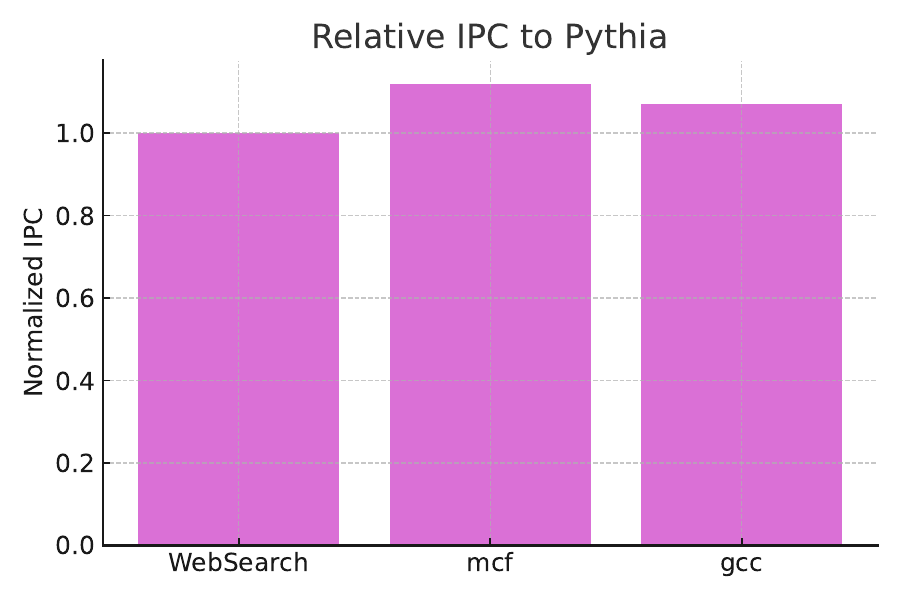}

\textbf{This figure presents the normalized IPC of CRL-Pythia relative to the baseline Pythia configuration. The IPC of baseline Pythia is set to 1.0 for each benchmark, and CRL-Pythia is plotted accordingly to visualize the relative performance improvement.}

\includegraphics[width=\columnwidth]{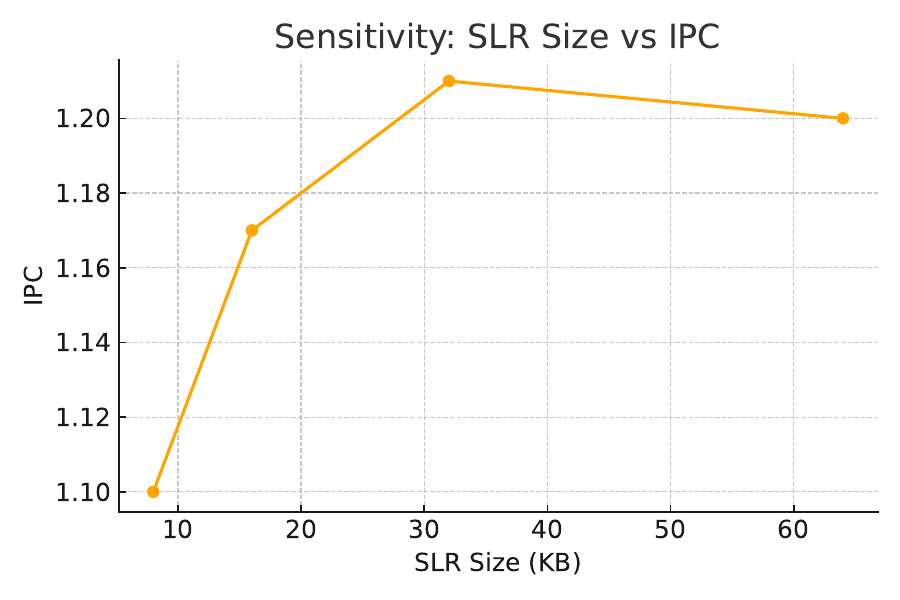}
\textbf{This line graph shows the sensitivity of system performance (IPC) to the size of the Shared Learning Repository (SLR). It evaluates the trade-off between learning effectiveness and memory overhead.}

\includegraphics[width=\columnwidth]{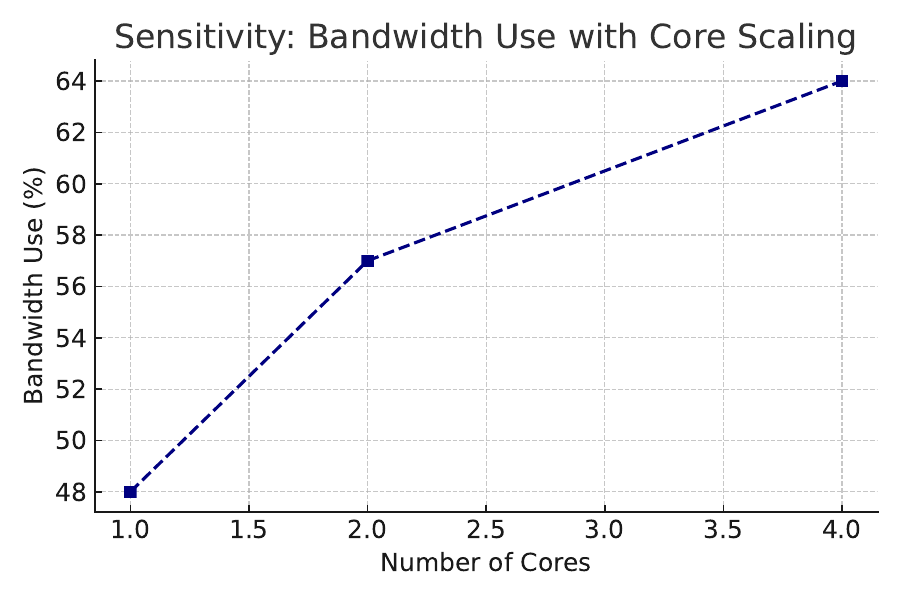}
\textbf{This chart depicts the change in memory bandwidth utilization as the number of active processor cores increases. It highlights how CRL-Pythia manages bandwidth more efficiently compared to baseline configurations.}

\includegraphics[width=\columnwidth]{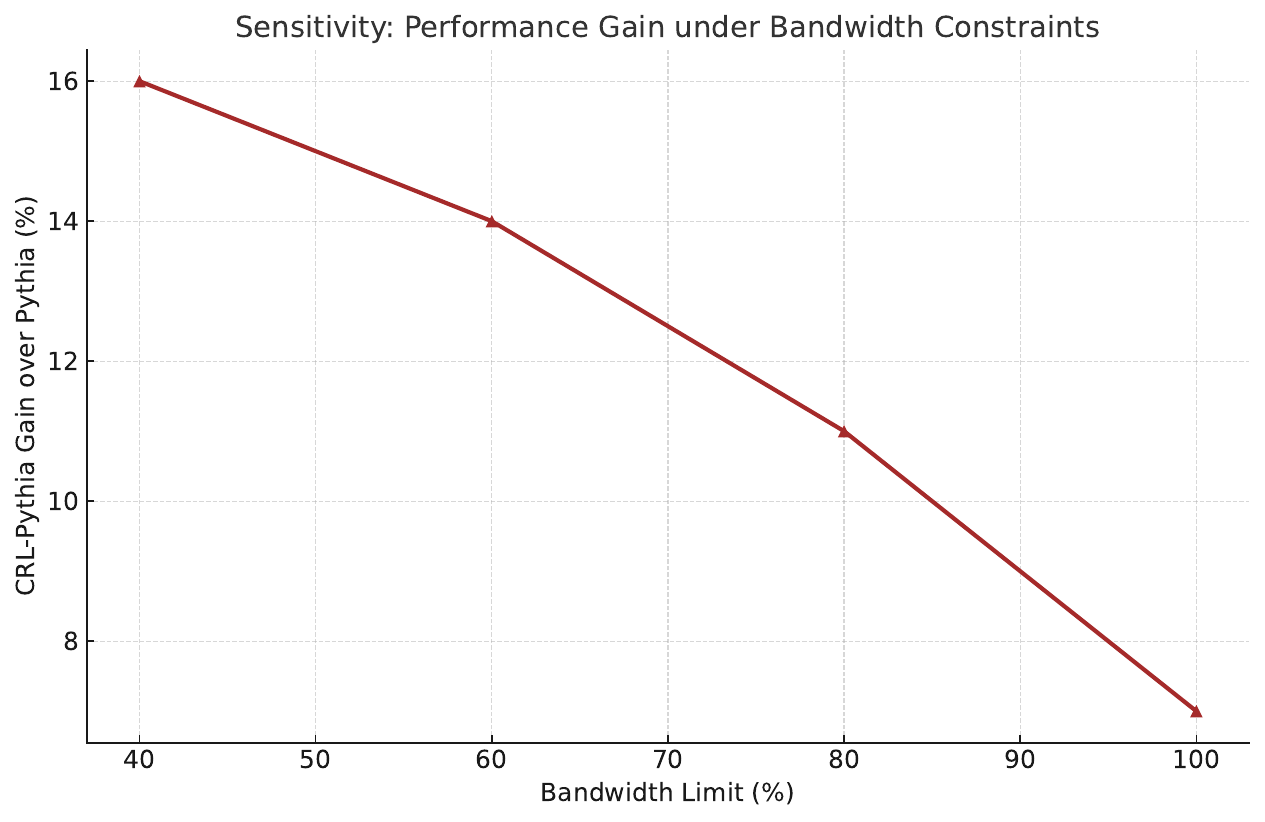}
\textbf{This figure illustrates the relative performance gain of CRL-Pythia over baseline Pythia under varying memory bandwidth constraints. It confirms that CRL-Pythia remains effective even under tight memory pressure.}

\section{Conclusion}
Our investigation identifies three key weaknesses in traditional prefetchers: limited feature exploitation, lack of system-level awareness, and rigid policies. Pythia alleviates these concerns using reinforcement learning, dynamically tailoring its behavior to workload characteristics. However, in highly regular patterns, methods like SPP+PPF occasionally outperform it, hinting at room for specialization in such contexts.
To further elevate multicore prefetching, we proposed CRL-Pythia, which embeds inter-core cooperation at the architectural level. By introducing the SLR and GST, our system enables collaborative learning and decision-making. Experiments show CRL-Pythia outpaces standalone prefetchers across varied workloads, particularly when bandwidth is constrained. Sensitivity analysis affirms its robustness and adaptability across core counts and system settings.
Future work includes: (1) integrating specialized heuristics for stream-like patterns, (2) exploring more expressive RL models and encodings, and (3) dynamically adjusting SLR parameters based on runtime behavior. We also envision tighter hardware-software co-design to further reduce coordination costs. Our findings underscore the growing need for coordination-aware architectures in next-generation multicore systems.


\bibliographystyle{ACM-Reference-Format}
\bibliography{sample-references}

\end{document}